\documentclass[aps,prl,twocolumn,superscriptaddress]{revtex4}
\usepackage{amssymb}
\usepackage{graphicx}
\usepackage{amsmath}

\begin{document}

\title{Superradiant Phase Transition of Fermi Gases in a Cavity across a Feshbach Resonance}
\author{Yu Chen}
\author{Hui Zhai}
\author{Zhenhua Yu}
\email{huazhenyu2000@gmail.com}
\affiliation{Institute for Advanced Study, Tsinghua University, Beijing, 100084, P. R. China}

\begin{abstract}
In this letter we consider the superradiant phase transition of a two-component Fermi gas in a cavity across a Feshbach resonance. It is known that quantum statistics plays a crucial role for the superradiant phase transition in atomic gases; in contrast to bosons, in a Fermi gas this transition exhibits strong density dependence. We show that across a Feshbach resonance, while the two-component Fermi gas passes through the BEC-BCS crossover, the superradiant phase transition undergoes a corresponding crossover from a fermionic behavior on the weakly interacting BCS side, to a bosonic behavior on the molecular BEC side. This intricate statistics crossover makes the superradiance maximally enhanced either in the unitary regime for low densities, in the BCS regime for moderate densities close to Fermi surface nesting, or in the BEC regime for high densities. 
\end{abstract}

\maketitle

Recent experiment has combined atomic Bose-Einstein condensates and cavity quantum electrodynamics together where atom-light interactions are strongly enhanced \cite{Kimble, Esslinger0, Reichel}. A superradiant phase transition driven by external pumping field has been observed, across which atoms form a density-wave order  \cite{Black, Esslinger1, Esslinger2}, and roton mode softening has been found in the vicinity of this superradiant phase transition \cite{Esslinger_Roton}. Theoretical studies have extended to investigate noninteracting Fermi gases inside a cavity \cite{Larson,Meystre,Subir,Chen,Simons,Piazza}. It is shown that the Fermi statistics plays a crucial role in the superrandiant phase transition at moderate and high densities \cite{Chen,Simons,Piazza}. At moderate densities, Fermi surface displays a nesting structure and strongly enhances superradiance, when the momentum of the cavity light field matches the nesting momentum. While at high densities, Pauli blocking effect forbids a large part of atom-light scattering processes, and consequently, strongly suppresses superradiance. The strong density dependence marks the major difference between superradiances in ideal Fermi gases and Bose gases. 

During the past decade, another important development in cold atom physics is the study of strongly interacting two-component Fermi gases and the BEC-BCS crossover utilizing Feshbach resonance \cite{Giorgini,Chin}. The inter-atomic $s$-wave scattering length $a_\text{s}$ can be continuously changed by a Feshbach resonance, and the dimensionless parameter $-1/k_\text{F}a_\text{s}$ ($k_\text{F}$ is the Fermi momentum in the noninteracting limit) controls the BEC-BCS crossover. In the BCS limit of the crossover $-1/k_\text{F}a_\text{s}\rightarrow +\infty$, fermions form loosely bound Cooper pairs and the low-energy response is dominated by fermionic quasi-particles; the system recovers a noninteracting Fermi gas. In the BEC limit $-1/k_\text{F}a_\text{s}\rightarrow -\infty$, Cooper pairs transform into tightly bound bosonic molecules, and the system responses to external fields mainly as bosons. In between, when $a_\text{s}$ is so large that $-1/k_\text{F}a_\text{s}\approx 0$, the system is in a strongly interacting regime and its response shall exhibit both bosonic and fermionic characters. 

So far, Fermi gases with inter-atomic interactions in a cavity have been barely studied. In this work we consider a two-component Fermi gas in a cavity across a Feshbach resonance. Given that the gas can be continuously tuned between the fermion limit and the boson limit, and that atoms with different statistics have been shown to behave differently in the superradiant phase transition  \cite{Chen,Simons,Piazza}, the motivation of our study is to address how the statistics crossover manifests itself in the superradiant phase transition across a Feshbach resonance, and the physical consequence of this crossover. In experiments, the superradiant phase transition is usually driven by increasing the strength of pumping field. In this work we will reveal nontrivial dependence of the critical pumping strength on the density of fermions $n$ and the inter-atomic interaction strength characterized by $-1/k_\text{F}a_\text{s}$. Our results represent a manifestation of the interplay between strong interactions from Feshbach resonance and strong atom-light coupling in a cavity, and will provide insight for future experiments.

\begin{figure}
\includegraphics[width=6.0cm]{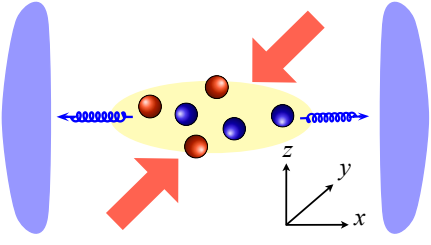}
\caption{ Experimental setup scheme. 
The pumping field propagates along the $\hat y$ direction shown
by the red arrows. The cavity field is represented by the wiggled lines in the $\hat x$ direction.
Fermions of different spins are shown in different colors.
}\label{setup}
\end{figure}

\emph{Model.}
Our system is schematically shown in Fig.~(\ref{setup}). Applied on the Fermi gas is a pumping field that consists of two laser beams counter-propagating along the $\hat{y}$ direction, with frequency $\omega_p$ and polarization in the $\hat z$ direction.
The single-mode cavity field of interest varies in the $\hat{x}$ direction, with frequency $\omega_c$ close to $\omega_p$.  The system is described by the Hamiltonian $\hat{\mathcal{H}}=\hat{\mathcal{H}}_\text{at}-\delta_c \hat{a}^\dag\hat{a}$, where $\hat a$ is the field operator for the cavity mode and $\delta_c=\omega_c-\omega_p$ is the cavity field detuning. 

The Hamiltonian experienced by the fermions has two parts $\hat{\mathcal{H}}_\text{at}=\hat{\mathcal{H}}_0+\hat{\mathcal{H}}_\text{int}$. The free fermion part is \cite{Chen}
\begin{align}{\label{eq:Hamil_BECBCS}}
& \hat{\mathcal{H}}_0=\sum_{\sigma=\uparrow,\downarrow}\int d^3{\bf r}\hat{\psi}_\sigma^\dag({\bf r})H_0\hat{\psi}_\sigma({\bf r}),\\
& H_0= \frac{{\bf p}^2}{2m}-\mu+V({\bf r})+ \eta({\bf r})(\hat{a}^\dag+\hat{a})+U({\bf r})\hat{a}^\dag \hat{a},\label{fh}
\end{align}
where $\hat{\psi}_\sigma({\bf r})$ are the fermion field operators with (peudo) spin index $\sigma\in\{\uparrow, \downarrow\}$. The pumping field and the cavity field generate respectively the optical potentials $V({\bf r})=V_0\cos^2(k_0y)$ and $U({\bf r})=U_0\cos^2(k_0x)$,  and the coupling between the pumping field and the cavity field comes from an interference term 
\begin{equation}
\eta({\bf r})=\eta_0\cos k_0x\cos k_0y
\end{equation}
with $\eta_0=\sqrt{U_0 V_0}$, $k_0$ is the wavevector magnitude of both the pumping field and the cavity mode \cite{supple}. The recoil energy $E_{\text{r}}= \hbar^2 k_0^2/2m$ is defined for latter use. 

The inter-atomic interaction nearby a Feshbach resonance is described by the Hamiltonian  
\begin{eqnarray}{\label{eq:Hamil_BECBCS1}}
 \hat{\mathcal{H}}_{\rm int}=g\int d^3{\bf r}\hat{\psi}^\dag_\uparrow({\bf r})\hat{\psi}^\dag_\downarrow({\bf r})\hat{\psi}_\downarrow({\bf r})\hat{\psi}_\downarrow({\bf r}).
\end{eqnarray}
The bare attractive inter-fermion interaction coupling $g$ is renormalized to the $s$-wave scattering length $a_s$ via $m/4\pi a_s=1/g+m\Lambda/2\pi^2$ with $\Lambda$ the momentum cutoff. This attractive interaction between fermions lead to fermion pairing and a Fermi superfluid ground state. 

\begin{figure}[b]
\includegraphics[width=5.5cm]{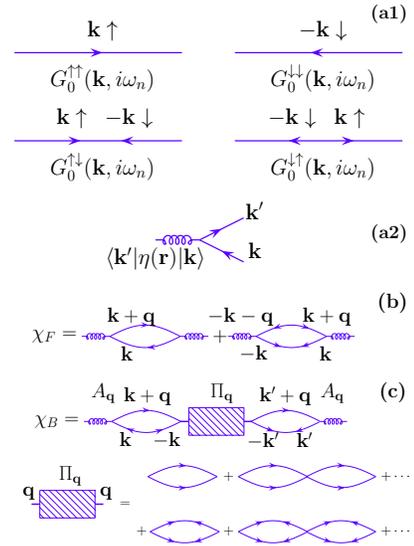}\\
\caption{(a1): the Feynman diagrams for propagators; the first line for the particle and hole propagators and the second line
for the anomalous Green's functions. (a2): the interaction vertex between the cavity field and the fermions.
(b,c), the Feynman diagrams corresponding to fermonic and bosonic contribution to density-wave susceptibility. The last line is the propagate of cooper pairs.  \label{diagram}}
\end{figure}

\emph{Ground State in Non-superradiant Phase.}
Before entering the suprradiance phase, $\langle \hat{a}^\dag\rangle=\langle \hat{a}\rangle=\langle \hat{a}^\dag\hat{a}\rangle=0$, fermions only experience a one-dimensional lattice $V({\bf r})$ along the direction of the pumping field, and the single-particle eigenstates are the Bloch states $|{\bf k}\rangle$ satisfying  $H_0|{\bf k}\rangle=\xi_{\bf k}|{\bf k}\rangle$. By expanding $\hat{\psi}_\sigma({\bf r})=\sqrt{1/V}\sum_{\bf k}\phi_{\bf k}({\bf r})\hat{c}_{\bf k\sigma}$ with $\langle {\bf r}|{\bf k}\rangle=\phi_{\bf k}({\bf r})$ and $V$ the gas volume, we introduce fermion pairing order parameter $\Delta_0={(g/V)}\sum_{\bf k}\langle c_{\bf k\uparrow}c_{-\bf k\downarrow}\rangle$$(\neq0)$. Here we assume the lattice $V({\bf r})$ is weak and we have ignored pairing at non-zero crystal momentum. With this assumption, the order parameter $\Delta_0$ is determined by the gap equation and the number equation \cite{supple}. In this Fermi superfluid state, the single-particle Green's functions are given by  
\begin{eqnarray}
G_0^{\uparrow\uparrow(\downarrow\downarrow)}({\bf k},i\omega_n)=-\frac{i\omega_n+(-)\xi_{\bf k}}{\omega_n^2+E_{\bf k}^2},
\end{eqnarray}
\begin{eqnarray}
G_0^{\uparrow\downarrow}({\bf k},i\omega_n)=G_0^{\downarrow\uparrow}({\bf k},i\omega_n)=\frac{\Delta_0}{\omega_n^2+E_{\bf k}^2},
\end{eqnarray}
with $E_{\mathbf k}=\sqrt{\xi_{\mathbf k}^2+\Delta_0^2}$ and the fermionic Matsubara frequencies $\omega_n=(2n+1)\pi/\beta$  for $n=0,\pm1,\pm2,\dots$, and $\beta$ the inverse of temperature. Their diagrams are shown in Fig.~(\ref{diagram})(a1). The components $G_0^{\uparrow\uparrow}(k)$ and $G_0^{\downarrow\downarrow}(k) (k\equiv({\bf k},i\omega_n))$ describe the propagation of particles and holes, respectively, while $G_0^{\uparrow\downarrow}(k)$ and $G_0^{\downarrow\uparrow}(k)$ are the anomalous Green's functions proportional to the pairing gap $\Delta_0$ which we take to be real. 

\emph{Condition for Superradiant Phase Transition.}
The superradiant phase transition is determined by the instability of non-superradiance toward developing non-zero $\langle \hat{a}^\dag \hat{a}\rangle$. As shown in Ref.~\cite{Esslinger1}, the superradiant phase transition occurs simultaneously with the formation of density-wave order of atoms with momentum ${\bf Q}_{\pm\pm}$, where  ${\bf Q}_{\pm\pm}=(\pm k_0,\pm k_0,0)$ is the momentum transfer between the cavity field and the pumping field. That is to say, $\langle\hat{a}\rangle$ is proportional to the density-wave order parameter  $\Theta=\int d^3{\bf r}\langle \hat n({\bf r})\rangle\eta({\bf r})/\eta_0$ with $\hat n({\bf r})=\sum_\sigma\psi^\dagger_\sigma({\bf r})\psi_\sigma({\bf r})$. By integrating out the fermion fields, one can obtain the free-energy of the system in the form $\mathcal{F}=\mathcal{C}\Theta^2$ \cite{supple}, where $\mathcal{C}$ changing sign from positive to negative gives the critical pumping field strength $\eta_0^{\rm cr}$ for the superradiance transition  \cite{Esslinger1,Chen,supple}
\begin{eqnarray}
\eta_0^{\rm cr}=\frac{1}{2}\sqrt{\frac{\tilde{\delta}_c^2+\kappa^2}{-\tilde{\delta}_c\chi}}\label{cri}.
\end{eqnarray}  
Here $\kappa$ is the cavity mode decay rate, and $\tilde{\delta}_c$ is the shifted cavity mode detuning $\tilde{\delta}_c=\delta_c-\int d^3\mathbf r \langle \hat n({\bf r})\rangle U(\mathbf r)$, which is assumed to be red-detuned ($\tilde{\delta}_c<0$). The most essential quality determining this transition is the density-wave order susceptibility of the Fermi superfluid defined as
\begin{align}
\chi=-\frac1{2\beta\eta_0^2}{\rm Tr}[
\langle T\hat n(\mathbf r_1,t_1)\hat n(\mathbf r_2,t_2) \rangle \eta(\mathbf r_1)\eta(\mathbf r_2)].
\end{align}
Here $\rm Tr$ includes the integration of the spatial coordinates and the imaginary times, $T$ is the time ordered operator. The expectation value of the fermion operators $\langle \dots\rangle$ is taken in the non-superradiant Fermi superfluid phase. A larger $\chi$ means that the Fermi gas has stronger tendency toward forming a density-wave order at a momentum ${\bf Q}_{\pm\pm}$, and it is easier for the Fermi gas to enter the superradiant phase; in another word, the critical pumping strength shall be smaller. 

\emph{Density-Wave Order Susceptibility.}
In order to capture both fermionic and bosonic responses of a Fermi superfluid, the density-wave order susceptibility $\chi$ should be calculated by the random phase approximation. This approximation maintains conservation laws \cite{baym, yu} and guarantees that one can recover the free fermion and the free boson results in the limits $-1/(k_\text{F}a_\text{s})\rightarrow \pm\infty$, respectively. Within this approximation, we have $\chi=\chi_\text{F}+\chi_\text{B}$ and

\begin{widetext}
\begin{eqnarray}
\!\!\chi_F\!&=&\!\!-\frac{1}{2\beta\eta_0^2}\!\sum_{{\bf k},{\bf k}',n}\!\!\!\!\left(G_0^{\uparrow\uparrow}({\bf k}',i\omega_n)G_0^{\uparrow\uparrow}({\bf k},i\omega_n)+G_0^{\downarrow\downarrow}({\bf k}',i\omega_n)G_0^{\downarrow\downarrow}({\bf k},i\omega_n)\right)|\langle{\bf k}|\eta({\bf \hat{r}})|{\bf k}'\rangle|^2, \label{chiF}\\
\!\!\chi_B\!&=&\frac1V\!\!\sum_{\mathbf q={\bf Q}_{\pm\pm}}A_{\mathbf q}^*\Pi_{\mathbf q}A_{\mathbf q}, \label{chiB1}\\
A_{\bf q}&=&\!\!-\frac{1}{2\beta\eta_0}\sum_{{\bf k},{\bf k}',n}\left(G_0^{\uparrow\uparrow}({\bf k}',i\omega_n)
G_0^{\downarrow\uparrow}({\bf k},i\omega_n)-G_0^{\downarrow\downarrow}({\bf k}',i\omega_n)
G_0^{\uparrow\downarrow}({\bf k},i\omega_n)\right)\langle{\bf k}'|\eta({\bf \hat{r}})|{\bf k}\rangle\langle{\bf k}|\gamma_{\bf q}({\bf \hat{r}})|{\bf k}'\rangle, \label{chiB2}\\
\!\!\Pi^{-1}_{\mathbf q}\!&=&\!\!-\frac{1}{{g}}+\frac{1}{V\beta}\sum_{{\bf k},{\bf k}',n}\sum_{\bf q'={\bf Q}_{\pm\pm}}\left(
G_0^{\uparrow\uparrow}({\bf k}',i\omega_n)G_0^{\downarrow\downarrow}({\bf k},i\omega_n)+
G_0^{\uparrow\downarrow}({\bf k}',i\omega_n)G_0^{\downarrow\uparrow}({\bf k},i\omega_n)\right)
\!\!\langle{\bf k}'|\gamma_{\bf q}({\bf \hat{r}})|{\bf k}\rangle\!\langle{\bf k}|\gamma_{\mathbf q'}({\bf \hat{r}})|{\bf k}'\rangle, \label{chiB3}
\end{eqnarray}
\end{widetext}
where $\gamma_{\bf q}({\bf r})\!=\!\cos({\bf q}\cdot{\bf r})$ is the mode factor for Cooper pair fluctuations.

The fermionic response $\chi_\text{F}$ is due to that the cavity field couples to the fermonic excitations of the Fermi superfluid by breaking up Cooper pairs. The Feynman diagrams corresponding to $\chi_\text{F}$ are shown in Fig.~(\ref{diagram})(b). The diagrams describe the process that a fermion with momentum ${\bf k}$ is scattered to momentum ${\bf k}^\prime$, where the momentum transfer comes from the photon momentum change from the pumping field to the cavity field, as denoted by the vertex in Fig.~(\ref{diagram})(a2). Since all fermions are paired in the Fermi superfluid phase, this process must be accompanied by pair breaking. In the BCS limit where the pairing gap $\Delta_0$ is small and pairs are easy to break, $\chi_\text{F}$ is dominant in $\chi$ and could recover the transition for free fermions in the limit of vanishing pairing gap \cite{supple}. While in the BEC limit this process is strongly suppressed because of large pairing gap. 

The bosonic response $\chi_\text{B}$ originates from the process that the cavity field excites nonzero momentum Cooper pairs and corresponds to the diagram shown in Fig. ~(\ref{diagram})(c). In this process, one of the two fermions in the Cooper pair, say, the one with momentum ${\bf k}$, is scattered to momentum ${\bf k+q}$ by a photon. Thus, the Cooper pair acquires a finite momentum ${\bf q}$ and propagates with this fixed momentum ${\bf q}$ (up to a reciprocal lattice vector along $\hat{y}$). 
After another scattering with a photon, the Cooper pair returns to zero-momentum. Because of weak lattice $V(\mathbf r)$ we only take into account the contributions from the scattered Cooper pairs of momentum $|q_y|\le q_0$.
The Cooper pair propagator $\Pi_{{\bf q}}^{-1}$ is given in Eq.~(\ref{chiB3}) and its diagram in Fig.~(\ref{diagram})(c) which is a summation of ladder diagrams. There are two ways for a Cooper pair to propagate, through multiple scattering and through vacuum fluctuations, respectively. Both are included in Eq.~(\ref{chiB3}) and in the bottom of Fig.~(\ref{diagram})(c).  In the BEC limit $\chi_\text{B}$ is dominant in $\chi$ and $\chi_\text{B}\sim a_s\Delta_0^2\sim n$ recovering the free boson result. While in the BCS limit, $\chi_\text{B}\sim \Delta_0^2/k_F$ is exponentially suppressed \cite{supple}.

\begin{figure}
\includegraphics[width=7.0cm]{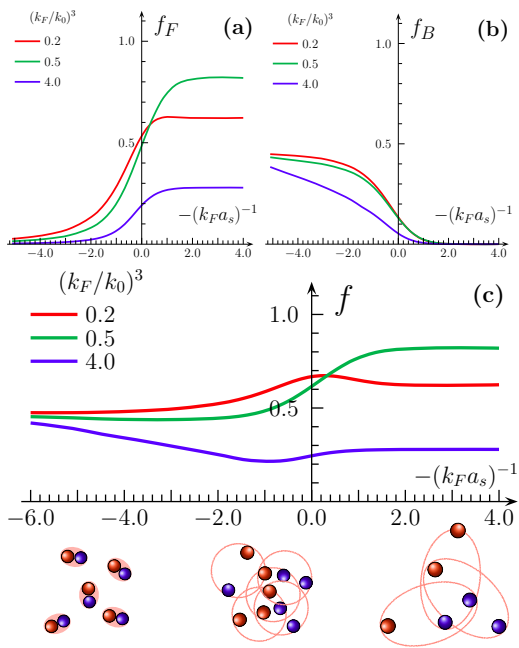}
\caption{Dimensionless susceptibilities $f_F$, $f_B$ and $f$ vs $-1/k_{\text F}a_{\text s}$ are plotted in (a), (b), (c) respectively with the pumping strength $V_0/E_r$ fixed at $0.1$ and $\nu$ taking $0.2$, $0.5$ and $4.0$. The bottom row is a pictorial representation of the BEC-BCS crossover.
 \label{fffbf}}
\end{figure}

We plot in Fig.~(\ref{fffbf}) the dimensionless susceptibility $f\equiv E_\text{r}\chi/N$, as well as its fermionic and bosonic constituent $f_F\equiv E_\text{r}\chi_F/N$ and $f_B\equiv E_\text{r}\chi_B/N$, as functions of the BEC-BCS crossover controlling parameter $-1/k_\text{F}a_\text{s}$, for different filling fractions $\nu=(k_F/k_0)^3$. 

First, Fig.~(\ref{fffbf})(a) shows that $f_\text{F}$ exhibits strong density dependence on the BCS side. Around a moderate density of $(k_\text{F}/k_0)^3=0.5$, the Fermi surface nesting is optimal, and $f_\text{F}$ becomes much larger than the low-density limit value $f_\text{F}=1/2$ \cite{Chen}. This is the regime where Fermi surface nesting strongly enhances superradiance, as discussed in noninteracting Fermi systems \cite{Chen,Simons,Piazza}. On the other hand, for high densities, say, $(k_\text{F}/k_0)^3=4$ in Fig.~(\ref{fffbf})(a), $f_\text{F}$ is much smaller than $1/2$ on the BCS side. This is the regime where the Pauli exclusion principle strongly suppresses superradiance. As approaching the BEC side, $f_\text{F}$ is strongly suppressed for all densities. Second, as shown in Fig.~(\ref{fffbf})(b), $f_\text{B}$ approaches the value of noninteracting bosons (also $=1/2$) in the BEC limit, independent of densities. While on the BCS side, for all densities $f_\text{B}$ is strongly suppressed.  

Figure~(\ref{fffbf})(c) shows the central result of this work. The total $f$ exhibits different features for different densities as $-1/k_{\text F}a_{\text s}$ varies. The most intriguing case is at relatively low-densities, say, $(k_\text{F}/k_0)^3=0.2$, where $f$ displays a maximum in the unitary regime ($1/a_\text{s}\approx0$). This maximum is because  in this regime, the bosonic contribution already takes off while the fermionic contribution has not damped out. While for moderate densities of Fermi surface nesting regime, $f$ monotonically increases as $-1/k_\text{F}a_\text{s}$ increases from the BEC limit to the BCS limit, due to the Fermi-surface nesting enhancement of $\chi$ on the BCS side. In contrast, for high densities, $f$ monotonically decreases, due to the Pauli blocking suppression of $\chi$ on the BCS side. The total $f$ has strong density dependence on the BCS side where it is dominated by the fermionic behavior, and becomes less and less sensitive to density in the BEC limit where it is dominated by the bosonic behavior. This change of $f$ with $-1/k_\text{F}a_\text{s}$ between the two limits is the manifestation of statistics crossover in superradiance.

\begin{figure}
\includegraphics[width=8.0cm]{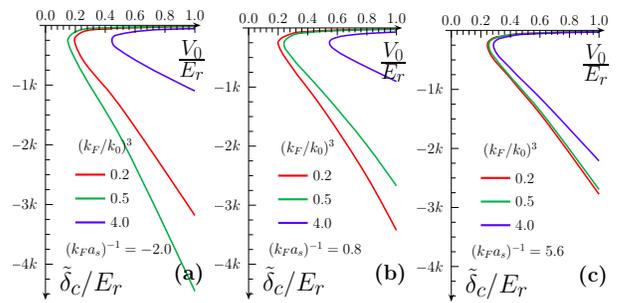}
\caption{Phase diagram for superradiance for different interaction parameter $1/k_{\text F}a_{\text s}=-2.0$ (a), $0.8$(b) and $5.6$ (c), and with different densities $\nu=0.2$, $0.5$ and $4.0$. For all cases, we take the typical experimental parameters
$\kappa/E_r=250$ and $U_0N_{\rm at}/E_r=10^3$.}\label{pd}
\end{figure}

\emph{Phase Diagram.} 
The boundary separating the normal and the superradiant phases can be obtained by solving Eq.~(\ref{cri}) \cite{supple}. In Fig.~(\ref{pd}), we plot the phase diagram in term of $V_0$ and $\tilde{\delta}$ for different densities and interaction strengths. In the BCS region, Fig.~(\ref{pd}) (a) shows that the moderate density $\nu=0.5$ is the easiest to be superradiant. In the unitary region as shown in Fig.~(\ref{pd}) (b) the low density $\nu=0.2$ is the easiest to be superradiant primarily due to the maximum of $f$ mentioned above in this part of the parameter space. 
On the BEC side, Fig.~(\ref{pd}) (c) shows that the density dependence diminishes since it shall be washed out completely in the BEC limit. 

\emph{Conclusion.} We have presented basic features of the superradiant phase transition of two-component Fermi gases across a Feshbach resonance. The main results are: i) On the BCS side of resonance the superradiant phase transition shows strong density dependence, similar as noninteracting Fermi gas; While on the BEC side it gradually becomes density independent, similar as noninteracting bosons. ii) Superradiance is mostly enhanced in the unitary regime for low density, in the BCS regime for moderate density, and in the BEC regime for high density.  
In this work, we have only focused on the superradiant phase transition itself. Inside the superradiant phase, the additional lattice due to the cavity field will further modify the single-particle dispersion, which will feedback to the Fermi superfluid. Furthermore, the quantum fluctuation of the cavity field will also generate additional effect on the Fermi superfluid. The properties of Fermi superfluids in the superradiant phase would be a subject for future studies. 

\emph{Acknowledgements.} This work is supported by Tsinghua University Initiative Scientific Research Program, NSFC under Grant No.~11004118, No.~11174176, No.~11104157, No.~11474179 and No.~11204152, and NKBRSFC under Grant No. 2011CB921500.

\newpage

\section{Supplementary Material}

\textbf{Model Parameters.}
The single fermion Hamiltonian Eq.~(\ref{fh}) is obtained by adiabatically integrating out all the electronic excitation states of the atoms in the rotating wave frame. The parameters in Eq.~(\ref{fh}) are related to the experimental tunable parameters as
$V_0=\Omega_p^2/\delta_a$, $U_0=g_0^2/\delta_a$ and $\eta_0=\sqrt{U_0 V_0}=\Omega_p g_0/\delta_a$. Here $\Omega_p$ is the pumping field strength, $\delta_a$ is the pumping laser frequency detuning with respect to electronic transitions of atoms, and $g_0$ is the coupling strength between the cavity mode and the fermions.

\vspace{0.1in}

\textbf{Mean Field Equation for Fermi Superfluids.}
When the lattice induced by the pumping field is not strong, we can approximate $\Delta({\bf r})={g}\langle\hat{\psi}({\bf r})\hat{\psi}({\bf r})\rangle=\Delta_0$ as a constant. The mean field gap equation becomes
\begin{eqnarray}
\Delta_0=\frac{{g}}{\beta V}\sum_{{\bf k},i\omega_n}G^{\uparrow\downarrow}({\bf k},i\omega_n)=-\frac{1}{V}\sum_{\bf k}\frac{{g}\Delta_0}{2E_{\bf k}}.
\end{eqnarray}
Together with the number equation $n=\frac{1}{V}\sum_{\bf k\sigma}\langle c_{\bf k\sigma}^\dag c_{\bf k\sigma}\rangle=\frac{1}{\beta V}\sum_{\bf{k},i\omega_n}(G^{\uparrow\uparrow}({\bf k},i\omega_n)-G^{\downarrow\downarrow}({\bf k},i\omega_n))$, or more explicitly,
\begin{eqnarray}
n=\frac{1}{V}\sum_{\bf k}\left(1-\frac{\xi_{\bf k}}{E_{\bf k}}\right),
\end{eqnarray}
we can determine $\Delta_0$ and $\mu$ self-consistently for a given pumping strength $V_0/E_r$ and given density $n$.

\vspace{0.1in}

\textbf{Instability Condition for Superradiant Phase Transition.} The mean field value of the cavity field $\alpha=\langle \hat a\rangle$ satisfies \cite{Esslinger1}
\begin{eqnarray}
i\frac{\partial\alpha}{\partial t}=(-\tilde{\delta}_c-i\kappa)\alpha+\eta_0\Theta,
\end{eqnarray}
where $\Theta=\int d^3{\bf r}\langle \hat n({\bf r})\rangle\eta({\bf r})/\eta_0$ is the fermion density order parameter. The introduced decay rate $\kappa$ is to model the weak leakage of electromagnetic field from the high-\emph{Q} cavity. 
In a steady state, $\partial_t\alpha=0$; we have 
\begin{eqnarray}\label{eq:PhaseLock}
\alpha=\frac{\eta_0\Theta}{\tilde{\delta}_c+i\kappa}, \label{mean_alpha}
\end{eqnarray}
which locks the cavity field to the fermion density order parameter. Both $\alpha$ and $\Theta$ are zero in the normal phase and become nonzero in the superradiant phase.

To the second order of $\alpha$, the effective free energy can be obtained as
\begin{eqnarray}
F_\alpha=-\frac{1}{\beta}\ln {\cal Z}_\alpha=-\tilde{\delta}_c\alpha^*\alpha-\chi(\alpha^*+\alpha)^2,\label{fa}
\end{eqnarray}
where ${\cal Z}_\alpha={\rm Tr}e^{-\beta H}$ with a specified $\alpha$.
By substituting (\ref{mean_alpha}) into Eq.~(\ref{fa}), we have
\begin{eqnarray}
F=-\left[\frac{\tilde{\delta}_c}{\tilde{\delta}_c^2+\kappa^2}+\chi
\frac{4\tilde{\delta}_c^2\eta_0^2}{(\tilde{\delta}_c^2+\kappa^2)^2}\right]\eta_0^2\Theta^2,\label{freeenergy}
\end{eqnarray}
which determines the superradiant transition when the quadratic coefficient of $\Theta$ changes its sign.

\vspace{0.1in}

\textbf{Explicit Expression for Density-Wave Order Susceptibility.} The explicit expressions for the density-wave order susceptibility within the BCS theory are 
\begin{align}
\chi_F=&\sum_{{\bf k},{\bf k}'}\frac{\left|\langle{\bf k}'|\eta({\bf \hat{r}})|{\bf k}\rangle\right|^2}{2\eta_0^2(E_{\bf k}+E_{{\bf k}'})}\left(1-\frac{\xi_{\bf k}\xi_{{\bf k}'}-\Delta_0^2}{E_{\bf k}E_{{\bf k}'}}\right),\label{chi33e}\\
A_{{\bf q}}=&\sum_{{\bf k},{\bf k}'}\frac{\langle {\bf k}'|\eta({\bf \hat{r}})|{\bf k}\rangle\langle {\bf k}|\gamma_{\bf q}(\bf \hat{r})
|{\bf k}'\rangle
}{2\eta_0(E_{\bf k}+E_{{\bf k}'})}
\frac{\Delta_0(\xi_{\bf k}+\xi_{{\bf k}'})}{E_{\bf k}E_{{\bf k}'}},\\
\Pi_{{\bf q}}^{-1}=&-\frac{V}{g}+\sum_{{\bf k},{\bf k}'}\sum_{\bf q={\bf Q}_{\pm\pm}}
\frac{2\langle {\bf k}'|\gamma_{\bf q}(\bf \hat{r})|{\bf k}\rangle \langle {\bf k}|\gamma_{\bf q'}(\bf \hat{r})|{\bf k}'\rangle
}{2(E_{\bf k}+E_{{\bf k}'})}\nonumber\\
&\times\left(1+\frac{\xi_{\bf k}\xi_{{\bf k}'}-\Delta_0^2}{E_{\bf k}E_{{\bf k}'}}\right).
\end{align}
In the BSC limit, the factor $1-\xi_{\bf k}\xi_{{\bf k}'}/E_{\bf k}E_{\bf k'}\approx n_F(\xi_{\bf k})-n_F(\xi_{\bf k'})$ with $n_F$ the Fermi-Dirac distribution; $\chi_F$ becomes the same as it is for free fermions \cite{Chen}. In the BEC limit, $A_{{\bf q}}\approx m^2a_s \Delta_0V\delta_{\mathbf q, \bf Q_{\pm,\pm}}/16\pi$ and 
$\Pi_{\bf q}\approx-16\pi/k_0^2ma_s$,  $f\approx1/2$ which is the same as it is for condensed noninteracting bosons \cite{Esslinger1, Chen}.

\vspace{0.1in}

\textbf{Determination of Phase Boundary.}
The boundary between the non-superradiant and superradiant phases is determined by Eq.~(\ref{cri}). Since $\eta_0=\sqrt{U_0V_0}$, $\chi=Nf/E_r$, and $f$ is a dimensionless function of dimensionless parameters $V_0/E_r$ and
$\nu=(k_F/k_0)^3$, we could recast Eq.~(\ref{cri}) in the form
\begin{align}
\frac{V_0}{E_r}\frac{NU_0}{E_r}f\left(\nu,\frac{V_0}{E_r}\right)=\frac{x^2+(\kappa/E_r)^2}{-x}
\end{align}
by introducing $x=\tilde{\delta}_c/E_r$. We take typical experimental values
$NU_0/E_r=10^3$ and $\kappa/E_r=250$. Thus at each
given pumping strength $V_0/E_r$ we can obtain the critical strengths of the cavity detuning $\tilde{\delta}_c$.

\end{document}